\documentclass{cargese}\usepackage{graphicx}

\let\footnote\savefootnote
\let\footnotetext\savefootnotetext 
\let\footnoterule\savefootnoterule 
\normallatexbib

\begin{document}


\articletitle[]
{Magnetized Type I Orbifolds \\
 in Four Dimensions}


\author{Marianna Larosa}



\affil{Dipartimento di Fisica, Universit\`a di Roma``Tor Vergata'', INFN, 
Sezione di Roma II \\
Via della Ricerca Scientifica 1, 00133 Roma, Italy}    

\email{Marianna.Larosa@roma2.infn.it
\\ ROM2F-02/31
}


\begin{abstract}
I  review the basic features of four dimensional $Z_2 \times Z_2$ (shift) 
orientifolds with internal magnetic fields, describing  two 
examples with $N=1$ supersymmetry. As in the corresponding six-dimensional 
examples, $D9$-branes magnetized along four internal directions can mimic 
 $D5$-branes, even in presence of 
multiplets of image branes localized on different fixed tori. Chiral
 low-energy spectra can be obtained if the model also contains
$D5$-branes parallel to the magnetized directions. 
\end{abstract}


In the last few years, open-string models have 
considerably enlarged our view of consistent string vacua and have also 
acquired a  valuable role for low-energy phenomenology. 
This is related to the presence  of
D-branes, where gauge and matter fields can be localized while 
gravity pervades the whole higher-dimensional bulk.
Type I models can be described as orientifolds of the Type IIB string theory 
by the world-sheet parity transformation $\Omega$ \cite{cargese}, and
compactifications  on various toroidal orbifolds have also allowed for 
attractive brane supersymmetry (breaking) configurations 
\cite{Antoniadis:1998ki,bsb}. 
Two kinds of exactly solvable
deformations can be introduced in Type I orbifold
models: (continuous and discrete) Wilson lines and internal
magnetic fields. Continous Wilson lines shift 
the momenta of open string states 
according to their charges, thus leading to gauge symmetry breaking 
while preserving the overall rank.
In a T-dual picture, this effect  can be ascribed to D-brane displacements 
in the ambient-space. On the other hand, discrete deformations (\emph{e.g.}
a quantized $B_{ab}$) lead in general to gauge group rank reductions.
The effect of uniform magnetic fields $H_i$ introduced in compact 
internal space is also relatively simple, since they couple only 
to the string endpoints that, say, carry charges $q_L$ and $q_R$.
The result, however, is  different according to 
whether the total  charge 
$Q=q_L+q_R$ is equal to or different from zero \cite{mf}.
 In the latter case the oscillator frequencies
 are shifted,  and there is a degeneracy in the 
Landau levels related to the non-commutativity of the string zero-modes.
 On the other hand, in the neutral case  the 
oscillator modes are unshifted, but the zero-modes are affected, and involve 
rescaled (boosted) contributions
to the one-loop annulus partition function. These
magnetic deformations admit  an alternative  interpretation in 
terms of rotated branes \cite{Berkooz:1996km}. In these 
types of constructions, related to magnetized branes by T-dualities, and
 widely analyzed in recent attempts to recover the Standard Model as a 
low-energy limit \cite{Blumenhagen:2001mb},
chiral fermions typically lie at D-brane intersections. 
Thus, in this contest the need for multiple matter families 
translates into the requirement of multiple D-brane intersections.
Furthermore, the introduction of magnetic fields
provides an interesting way to break 
supersymmetry in Type I models \cite{Bachas:1995ik}, for 
particles of different spins couple differently
to them via their magnetic moments, thus giving rise to different masses. 
For arbitrary configurations of the magnetic fields, 
the open spectrum is indeed non-supersymmetric, 
and tachyonic (Nielsen-Olesen) instabilities \cite{no} are present.
 However, for (anti)self-dual magnetic field configurations 
\cite{aads}, or, in the T-dual language, 
for particular  intersection angles 
between the D-branes \cite{Berkooz:1996km}, tachyons are absent and some
residual supersymmetry is preserved.

In the following I shall describe two classes of Type I 
\emph{magnetized} models in 
four dimensions, built  as open descendants 
of $Z_2 \times Z_2$ (shift) orbifolds. 
In both cases the starting point is the Type IIB string theory 
compactified on a six-dimensional torus $T^6 = T^2 \times T^2 \times T^2$,
and projected by the 
$Z_2 \times Z_2$ operations $o=(+,+,+), g=(+,-,-), h=(-,-,+)$ and
$f=(-,+,-)$, that, aside from the identity $o$, act as 
$\pi$-rotations on two of the three internal tori.
There are actually two classes of supersymmetric $Z_2 \times Z_2$ models, 
that differ because of the possible presence of discrete torsion, 
a sign $(\epsilon=\pm 1)$ in their partition functions,
with crucial effects both on the low-energy closed spectra and on the 
structure of the open descendants. Indeed, only 
models with discrete torsion ($\epsilon=-1$) give rise to 
open descendants containing chiral fermions, 
but chirality in the open spectrum can also be recovered, for $\epsilon=1$,
if suitable internal magnetic fields are introduced.
Let us  thus look more closely at  models without discrete torsion. 
They contain a set of 
$D9$-branes and three different sets of $D5$-branes, $D5_1, D5_2$
and $D5_3$. Homogeneous magnetic fields along two of the
three two-tori  alter the boundary conditions of the open strings stretched 
between these $D$-branes. 
Furthermore, in order to absorb the R-R charge 
of the background, the open sector contains also a set of \emph{magnetized} 
$D9$-branes that, as in the six-dimensional examples of 
\cite{aads}, mimic the behavior of one set of $D5$-branes. 
This is a \emph{brane transmutation}:
in  a magnetic field, a $D9$-brane can acquire a $D5$ RR charge 
\cite{Douglas:1995bn}, and 
can therefore contribute to the corresponding tadpole.
Chiral matter appears in the open spectrum when $D5$-branes extend
along the magnetized directions of the $D9$-branes, and typically 
arranges itself in multiple families. In a T-dual picture, the effect can 
be equivalently ascribed to the intersections
of mutually rotated $D7$-branes.
  
When the conventional orbifold operations are combined with momentum or 
winding shifts of the lattice states, the resulting models can display a 
partial breaking of supersymmetry, recovered for large or 
small radii, with  peculiar 
$D5$-brane configurations, that typically arrange themselves in multiplets of
images, interchanged by some of the orbifold operations. This is what happens,
for instance, for the $w_2p_3$ models  of \cite{shift},
with shifts of the lattice windings and 
momenta in the second and third internal tori.
They have one set of $D9$-branes and two sets of $D5$-branes, $D5_1$ and
 $D5_2$, that  appear organized in doublets and four-plets.
The introduction of two magnetic fields in the second and third tori gives 
rise to a model with a chiral spectrum, as expected from the presence of the 
$D5_2$ branes, that extend along a magnetized compact direction.
Moreover, the \emph{brane transmutation} phenomenon appears in 
this case in a subtler way, again because of the presence of 
$D5$ multiplets. 
These  are distributed among the orbifold fixed points, and
thus it is not possible to insert all branes at a single fixed point. 
The $D9$-branes do their  best to mimic the same behavior, even though
they are not localized, owing to their spatial extension. 
However, the centers of the classical Landau orbits organize themselves 
in doublets as well, so that  each orbit is  accompanied
 by images, just like each $D5$-brane in the undeformed model.
Even in this case the open spectra reveal the presence of 
multiplets of matter  families and an unusual rank reduction of the 
corresponding Chan-Paton group. An interesting by-product-of this analysis
is that the familiar discrete deformations, induced by a quantized $B_{ab}$,
can also be generated by suitable shift-orbifolds of this type.
A more detailed discussion of these models may be found in 
\cite{Larosa:2001ku, noi}.
\begin{acknowledgments}
It is a pleasure to thank G. Pradisi for a stimulating  
collaboration and A. Sagnotti for introducing 
me to these topics. 
I am also grateful to the Organizers of the Carg\`ese 2002 ASI
for giving me the opportunity to present these results.
This work was supported in part by I.N.F.N., by the
E.C. RTN programs HPRN-CT-2000-00122 and HPRN-CT-2000-00148, by the 
INTAS contract 99-1-590, by the MURST-COFIN contract 2001-025492 and 
by the NATO contract PST.CLG.978785.

\end{acknowledgments}


\end{document}